# Tunable Topological Phase Transitions in a Piezoelectric Janus Monolayer

Tanshia Tahreen Tanisha[1], Md. Shafayat Hossain[2*], Nishat Tasnim Hiramony[1], Ashiqur Rasul[1], M. Zahid Hasan[2*], Quazi D. M. Khosru[1*]

[1]Department of Electrical and Electronic Engineering, Bangladesh University of Engineering and Technology, Dhaka, 1205, Bangladesh

[2]Department of Physics, Princeton University, Princeton, NJ, 08544, USA

*Corresponding authors. E-mails: mdsh@princeton.edu; mzhasan@princeton.edu; qdmkhosru@eee.buet.ac.bd.

**ABSTRACT:**

Quantum Spin Hall (QSH) insulators represent a quintessential example of a topological phase of matter, characterized by a conducting edge mode existing within a bulk energy gap. The pursuit of a tunable QSH state stands as a pivotal objective in the development of QSH-based topological devices. In this study, we employ first-principles calculations to identify three strain-tunable quantum spin Hall (QSH) insulators based on monolayer $MAlGaTe_4$ (where M represents Mg, Ca, or Sr). These monolayers exhibit dynamic stability, with no imaginary modes detected in their phonon dispersion. Additionally, they possess piezoelectric properties, rendering them amenable to strain-induced tuning. While MgAlGaTe$_4$ is a normal insulator under zero strain, it transitions into the QSH phase (hosting topologically protected, gapless edge states residing with a bulk bandgap) when subjected to external strain. Conversely, CaAlGaTe$_4$ and SrAlGaTe$_4$ already exhibit the QSH phase at zero strain. Intriguingly, upon the application of biaxial strain, these two compounds undergo phase transitions, encompassing metallic (M), normal/trivial insulator (NI), and topological insulator (TI) phases, thereby illustrating their strain-tunable electronic and topological properties. (Ca, Sr)AlGaTe$_4$, in particular, undergo M-TI/TI-M transitions under applied strain, while MgAlGaTe$_4$ additionally experiences an M-NI/NI-M transition, signifying it as a material featuring a metal-insulator transition (MIT). Remarkably, the observation of metal-trivial insulator-topological insulator transitions in MgAlGaTe$_4$ introduces it as a unique material platform in which both MIT and topological phase transitions can be controlled through the same physical parameter. Our study thus introduces a novel material platform distinguished by highly strain-tunable electronic and topological properties, offering promising prospects for the development of next-generation, low-power topological devices.

**Introduction:**

Going beyond the dichotomoy of metal and insulators, the topological insulators provides a conducting boundary mode within an insulating bulk.[1–4] Notably, this conducting boundary mode is free of backscattering and the time-reversal symmetry protected topological spectra flow is still maintained even in the presence of arbitrarily strong non-magnetic disorder. This property is particularly desirable for fabricating nanoelectronic devices. Quantum spin Hall (QSH) insulators are the two-dimensional variants of the topological insulators and are expected to be more convenient for developing applications. Indeed, various applications have been proposed, ranging from dissipationless conducting wires for nanoelectronic circuits to van der Waals topological field-effect transistors (vdW-TFET), based on the discovery of the 2D topological insulator phase in 1T' $MX_2$.[5] These materials are also considered as substrates to support catalysts[6] and hold potential for topological quantum computing applications.[7] Realizing these applications would be immensely benefitted from achieving reversible, electronic control over the QSH state. Here we predict a novel class of materials (Ca, Sr, Mg)AlGaTe$_4$ that facilitates such electronic control via the application of external strain.

Our proposition of the (Mg, Ca, Sr)AlGaTe$_4$ family of compounds is partly inspired by prior investigations into the septuple-atomic-layer $MA_2Z_4$ family,[8] which encompasses a diverse range of materials including direct-gap semiconductors, ferromagnetic semiconductors, topologically non-trivial insulators, and type-I Ising superconductors. Notably, MoSi$_2$N$_4$, WSi$_2$N$_4$, and MnBi$_2$Te$_4$, members of this family, have already been successfully synthesized.[9,10] Within this family, both SrAl$_2$Te$_4$ and SrGa$_2$Te$_4$ have been theorized to possess topologically non-trivial insulating properties. Consequently, it is plausible that their Janus counterpart, SrAlGaTe$_4$, may also exhibit such topological characteristics. On the contrary, while (Mg, Ca)Ga$_2$Te$_4$ are predicted to be topologically non-trivial, (Mg, Ca)Al$_2$Te$_4$ are anticipated to be topologically trivial. Hence, there exists a tantalizing possibility that the topological attributes of their Janus counterparts, MgAlGaTe$_4$ and CaAlGaTe$_4$, can be finely tuned, potentially leading to a transition from a trivial to a non-trivial state.

In the context of tuning the potential topological properties, strain engineering has emerged as a powerful tool for precisely controlling material properties, offering a promising avenue for tailoring electronic characteristics. Studies on methyl-functionalized germanene (GeCH$_3$)[11] and Silicene (SiCH$_3$)[12] serve as illustrative examples of the impact of strain on material behaviour. External uniaxial and biaxial strain, in particular, have been shown to effectively modulate the electronic bandgap of materials.[13–16] This capability holds significant potential for inducing metal-insulator phase transitions and vice versa, and for widening bandgaps to enable operation at higher temperatures.[17–21] These observations prompt us to explore the effects of strain on the (Mg, Ca, Sr)AlGaTe$_4$ family of compounds. An additional facet of these strain-dependent studies involves the investigation of piezoelectric properties, a characteristic exhibited by many Janus monolayers owing to their reduced symmetry.[22–25] Notably, the Janus Quantum Spin Hall (QSH) insulator, SrAlGaSe$_4$,[26] possesses piezoelectricity, and under biaxial strain, it undergoes a transition from a normal insulator to a piezoelectric QSH state. Similarly, the Janus monolayer VCClBr[27] exhibits piezoelectric QSH insulating properties but remains impervious to tuning via uniaxial or biaxial strain, or external electric fields. Nevertheless, the coexistence of the QSH effect and piezoelectricity in these materials positions them as valuable candidates for integration into multifunctional electronic and spintronic devices. In our exploration of (Mg, Ca, Sr)AlGaTe$_4$, we uncover the simultaneous presence of piezoelectricity, strain-tunable metal-insulator transitions, and the ability to reversibly shift from a topologically trivial state to a QSH state.

**Results and Discussions:**

The monolayer form of MAlGaTe$_4$ exhibits a hexagonal lattice structure. The top views (top panel) and side views (bottom panel) of the geometry-optimized structures of MgAlGaTe$_4$ and SrAlGaTe$_4$ are shown in **Fig. 1**. The MAlGaTe$_4$ unit cell comprises one Al atom, one Ga atom, four Te atoms, and one M atom, where M can be Mg, Ca, or Sr. This two-dimensional material constitutes a septuple layer system, signifying that the atoms are arranged across seven vertical layers. To gain a deeper understanding of the monolayer's structure, we focus on the side view of MgAlGaTe$_4$, as shown in **Fig. 1(a)**. This structure can be viewed as a composite arrangement, with a monolayer of MgTe$_2$ sandwiched approximately midway between an upper layer of AlTe and a lower layer of GaTe. Within the unit cell, four Te atoms play distinct roles: Te$_A$ is bonded to Al in the upper AlTe layer, Te$_D$ is bonded to Ga in the lower GaTe layer, and Te$_B$ and Te$_C$ are both bonded to Mg, constituting the middle MgTe$_2$ layer. Te$_B$ connects MgTe$_2$ with AlTe, while Te$_C$ bridges the gap between MgTe$_2$ and GaTe. This structural analysis applies similarly to CaAlGaTe$_4$ and SrAlGaTe$_4$. It is worth noting that, in most aspects, CaAlGaTe$_4$ and SrAlGaTe$_4$ exhibit similar properties, and the details regarding the former are provided in the **Electronic Supplementary Information (ESI)**. An interesting characteristic of these structures is their lack of centrosymmetry, distinct from their non-Janus counterparts, MAl$_2$Te$_4$ and MGa$_2$Te$_4$[8]. **Table 1** presents the geometry-optimized lattice parameters and bond lengths associated with the monolayers MAlGaTe$_4$, where M can be either Mg or Sr.

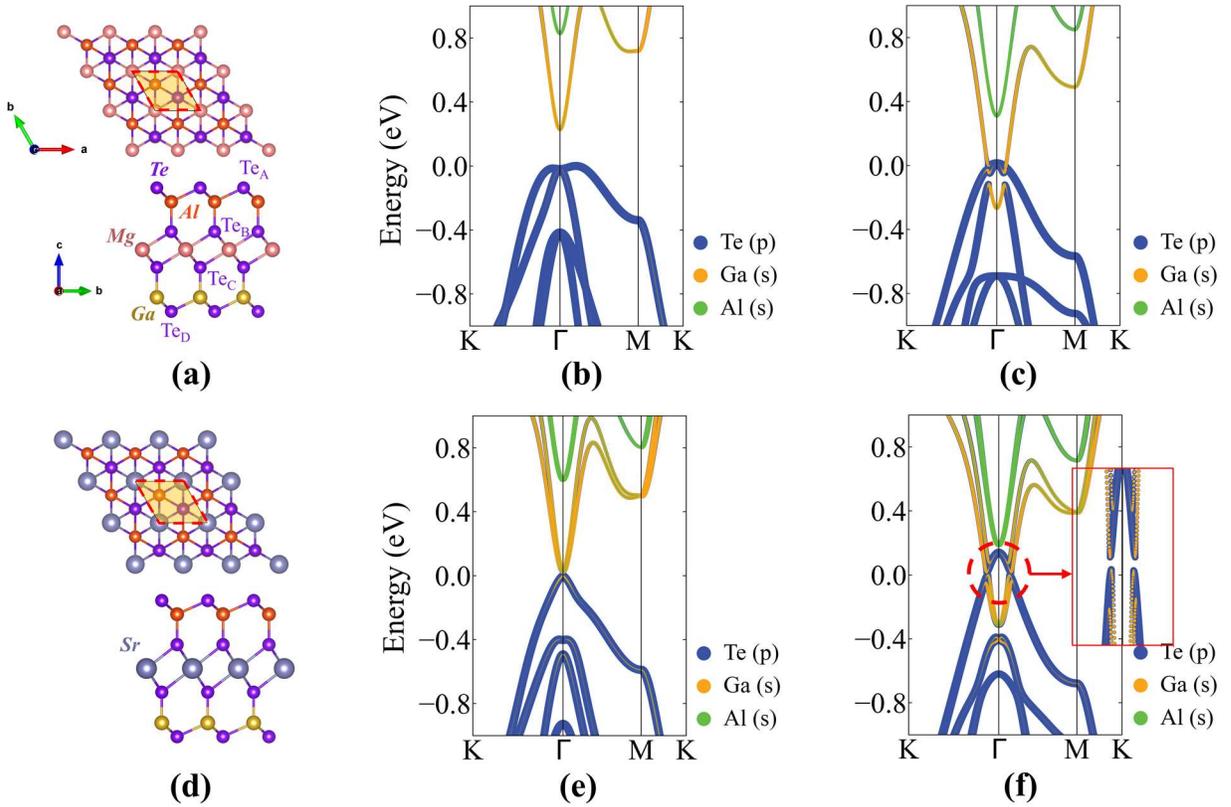

**Fig. 1. Lattice symmetry and electronic band structures of MgAlGaTe$_4$ and SrAlGaTe$_4$.** **(a)** and **(d)**: Top-view (above) and side-view (below) representations of MgAlGaTe$_4$ and SrAlGaTe$_4$, respectively. In the top-views, the unit cells are delineated by semi-transparent orange parallelograms. **(b)** and **(c)**: Orbital-resolved band structures without Spin-Orbit Coupling (SOC) for MgAlGaTe$_4$ and SrAlGaTe$_4$, respectively. **(e)** and **(f)**: Corresponding band structures in the same order, accounting for the inclusion of SOC. The inset in **(f)** provides a magnified view of the inverted bandgap observed in SrAlGaTe$_4$.

**Table 1. Lattice constants and bond lengths of MgAlGaTe$_4$ and SrAlGaTe$_4$.** $d_1$-$d_6$ represent the bond lengths corresponding to Al-Te$_A$, Al-Te$_B$, M-Te$_B$, M-Te$_C$, Ga-Te$_C$, and Ga-Te$_D$ bonds, respectively. The bonds are visually represented and labeled in **Fig. S2 (a)** of the **ESI.**

| Material | Lattice constant (Å) | $d_1$ (Å) | $d_2$ (Å) | $d_3$ (Å) | $d_4$ (Å) | $d_5$ (Å) | $d_6$ (Å) |
|---|---|---|---|---|---|---|---|
| MgAlGaTe$_4$ | 4.27 | 2.74 | 2.55 | 2.91 | 2.89 | 2.57 | 2.75 |
| SrAlGaTe$_4$ | 4.39 | 2.78 | 2.52 | 3.22 | 3.21 | 2.55 | 2.78 |

First, we verify the dynamic and mechanical stability of the proposed materials. To achieve this, we conduct thorough calculations of the phonon dispersion spectra and the elastic constants tensors ($C_{kj}$) for the three monolayers (detailed in sections A and B of the **ESI**). Our findings confirm the stability of these monolayers, as they exhibit no imaginary modes in their phonon dispersion.

The orbital-resolved electronic band structures of MAlGaTe$_4$ (M: Mg, Sr) with and without spin-orbit coupling (SOC) are presented in **Fig. 1**, computed using GGA functionals. In the case of monolayer MgAlGaTe$_4$, the absence of SOC yields an indirect bandgap of 229.6 meV (**Fig. 1(b)**). Upon introducing SOC, the bandgap becomes direct, with a reduced value of 23.3 meV (**Fig. 1(e)**). Notably, neither band structure displays a band inversion, indicating that monolayer MgAlGaTe$_4$ is topologically trivial. On the other hand, the band structures of SrAlGaTe$_4$ reveal a different behaviour. In the absence of SOC, no bandgap is evident (**Fig. 1(c)**). However, upon including SOC, an inverted bandgap of 14.5 meV emerges (**Fig. 1(f)**). Here, an electronic band inversion is discernible near the valence band maximum (VBM) and conduction band minimum (CBM), a characteristic signature of topologically non-trivial insulators. The inset provides a magnified view of the inverted bandgap obtained with SOC. CaAlGaTe$_4$ exhibits similar electronic traits, and detailed band structures are provided in section C of **ESI**.

At zero strain, MgAlGaTe$_4$ exhibits no band inversion. It is worth noting that, MgAl$_2$Te$_4$ possesses $\mathbb{Z}_2 = 0$, while MgGa$_2$Te$_4$ has $\mathbb{Z}_2 = 1$. Therefore, it is plausible that their Janus compound, MgAlGaTe$_4$, may demonstrate tunable topological properties. To investigate this, we introduced biaxial strain on MgAlGaTe$_4$ by adjusting the lattice constants using the formula: $\epsilon = \frac{a-a_0}{a_0} \times 100\%$. Here, $\epsilon$ represents the percentage strain, $a_0$ is the lattice constant of the unstrained, relaxed structure, and $a$ is the lattice constant of the strained structure.

The orbital-resolved band structures of MgAlGaTe$_4$ without and with SOC under various strains (-4%, -2%, 0%, 2%, and +4%) are depicted in **Fig. 2**. In the presence of SOC and without strain, the VBM is predominantly occupied by the p-orbitals of Te atoms, while the CBM is primarily populated by s-orbitals of Ga atoms (**Fig. 2(h)**). At compressive strains of 2% and 4%, no band inversion occurs (**Figs. 2(f), (g)**). However, at 2% tensile strain (**Fig. 2(i)**), a band inversion emerges, signifying an interchange in the dominant orbital characteristics of the CBM and VBM. Notably, the VBM now exhibits a significant occupation of the s-orbitals of Ga atoms, while the CBM displays a substantial percentage of p-orbitals of Te atoms. At this strain, a small bandgap of 17 meV materializes. This band inversion persists at 4% tensile strain (**Fig. 2(j)**) and the inverted bandgap widens further to 39 meV. This observed band inversion offers the initial indication of a QSH state in MgAlGaTe$_4$ upon the application of biaxial tensile strain.

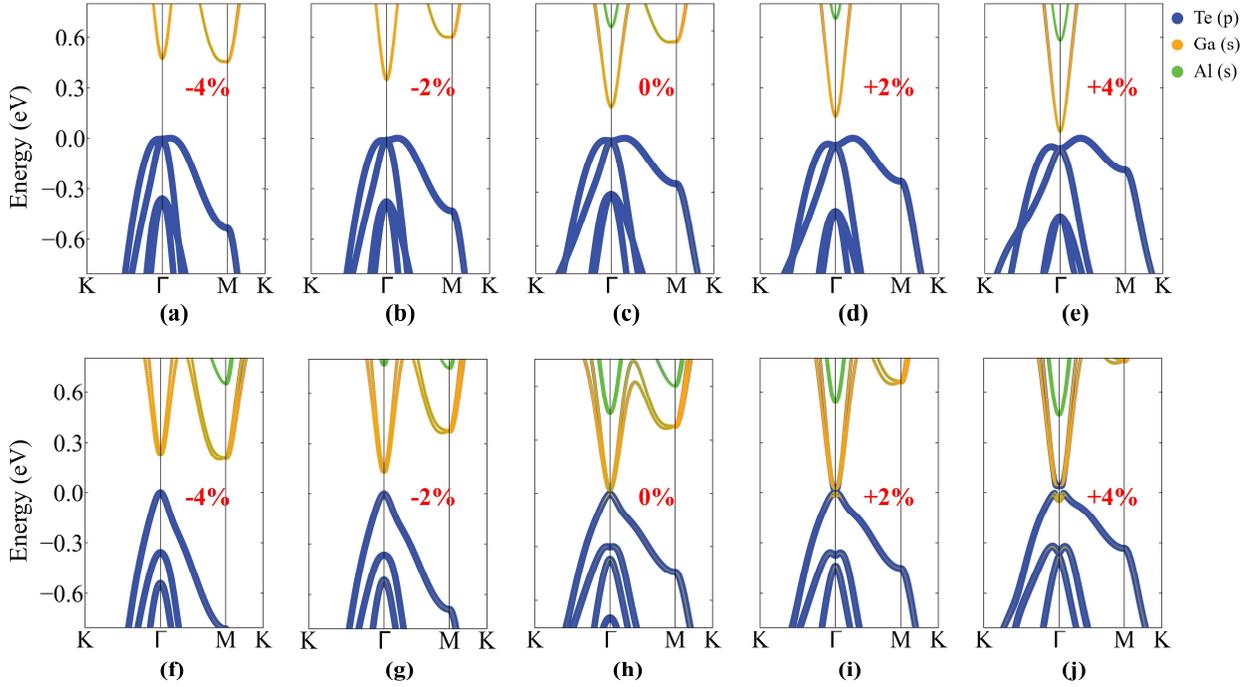

**Fig. 2. Orbital-resolved bandstructures of monolayer MgAlGaTe$_4$ under various strain conditions.** The strain values are: **(a, f)** -4%, **(b, g)** -2%, **(c, h)** 0%, **(d, i)** +2%, and **(e, j)** +4%. Panels **(a-e)** and **(f-j)** present calculation results without and with the incorporation of SOC, respectively.

To confirm the presence of the QSH effect in strained MgAlGaTe$_4$, we conducted calculations of the edge density of states and the $\mathbb{Z}_2$ topological invariant. These calculations involved the intersections of the Wannier Charge Center (WCC) spectra with a randomly positioned horizontal reference line. For MgAlGaTe$_4$, under 2% biaxial tensile strain, the calculations reveal $\mathbb{Z}_2 = 0$, and hence, these results are not presented here. Subsequently, we repeated the calculations at 4% biaxial tensile strain, and the results are illustrated in **Fig. 3**. The WCC spectra, as depicted in **Fig. 3(a)**, intersect the reference line only once (indicating an odd number of intersections). This observation suggests $\mathbb{Z}_2 = 1$ for this particular case. The corresponding edge density of states, shown in **Fig. 3(b)**, clearly demonstrates the presence of edge states connecting the conduction and valence bands. These findings, including the band inversion, the existence of edge states, and the non-zero topological invariant $\mathbb{Z}_2 = 1$, unequivocally confirm the existence of the QSH effect in MgAlGaTe$_4$ under 4% biaxial tensile strain.

We have also computed the edge density of states and $\mathbb{Z}_2$ invariant in the unstrained SrAlGaTe$_4$, as depicted in **Figs. 3(c)** and **3(d)**. Akin to the findings in MgAlGaTe$_4$ under 4% biaxial tensile strain, the WCC spectra intersects the red dashed horizontal reference line once, indicating $\mathbb{Z}_2 = 1$. Additionally, the edge density of states demonstrates the presence of gapless edge states, forming connections between the gapped bulk conduction and valence bands. This observation serves as compelling evidence that SrAlGaTe$_4$, even at zero strain, already harbours the QSH effect. Similarly, CaAlGaTe$_4$ exhibits analogous topological properties, and the corresponding edge density of states and $\mathbb{Z}_2$ invariant are presented in section D of the **ESI**.

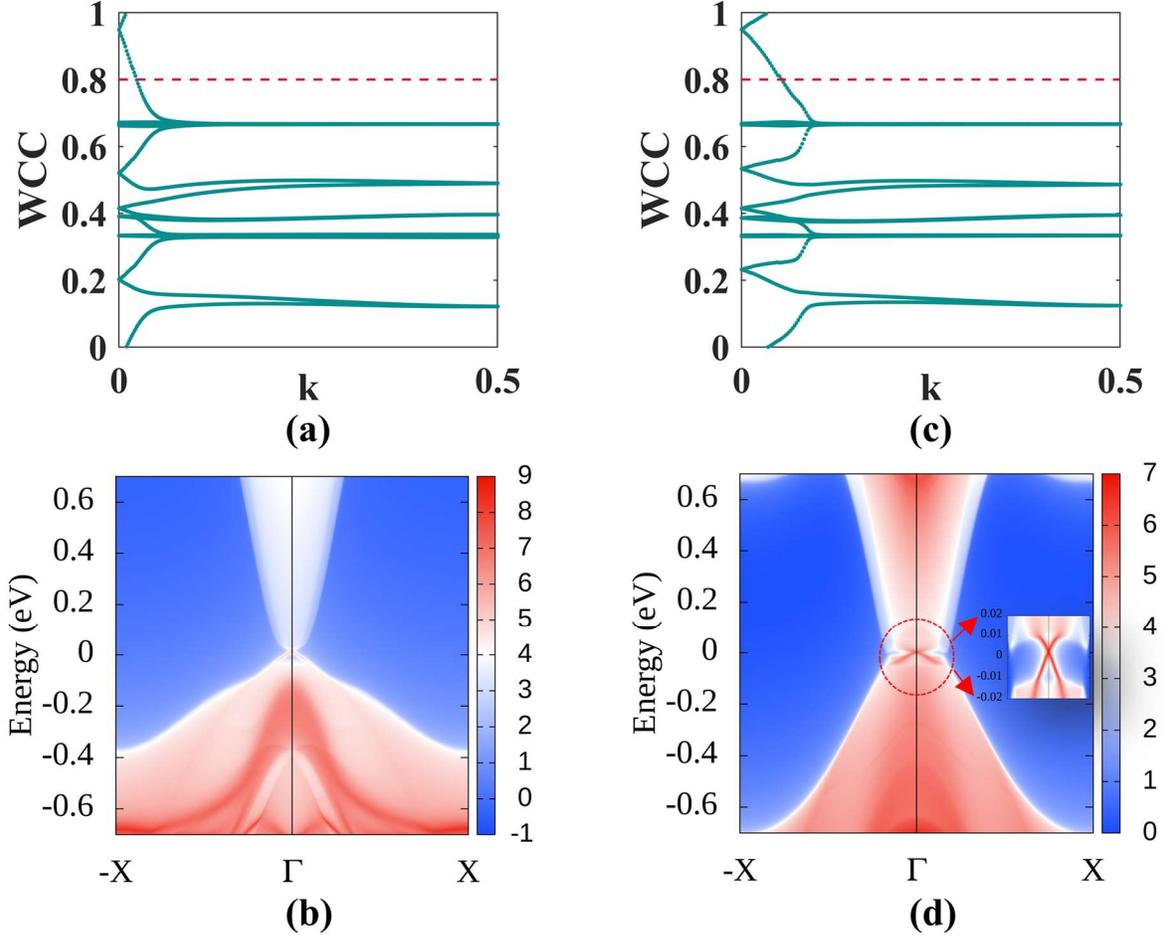

**Fig. 3. Topological invariant and in-gap edge modes. (a), (c)** depict the WCC spectra of MgAlGaTe$_4$ (at 4% biaxial tensile strain) and SrAlGaTe$_4$, respectively. The WCC spectra intersect the horizontal reference line once (odd), indicating $\mathbb{Z}_2 = 1$. **(b), (d)** display the corresponding edge density of states in the same order, featuring gapless edge states. The inset in **(d)** provides a magnified view of the edge states for SrAlGaTe$_4$.

Revisitng the data in **Fig. 2**, which presents the electronic band structure under biaxial strain ranging from -4% to +4% for MgAlGaTe$_4$ with 2% intervals, we find that the external strain significantly influences the bandgap of MgAlGaTe$_4$. Moreover, we have extended our analysis to include bandstructures for ±6% and ±8% strain, presented in **Fig. S4** of the **ESI**. Specifically, at - 6% and -8% strain, the CBM resides in the M point and crosses the VBM, leading to a scenario where MgAlGaTe$_4$ displays no bandgap and behaves as a metal. However, at -4% strain, the CBM and VBM separate in energy, giving rise to a large indirect bandgap, marking a metal-insulator transition (MIT) between -6% and -4% strain. As we progress to -2% strain, the CBM shifts to the Γ point, forming a direct bandgap; nevertheless, it remains a trivial insulator, as discussed previously. With a further change in strain to +2%, an inverted bandgap materializes. Ultimately, at +4% strain, the QSH state emerges, pointing to a transition from a trivial insulator to a topological insulator. The QSH state persists at +6% and +8% strain, retaining the band inversion.

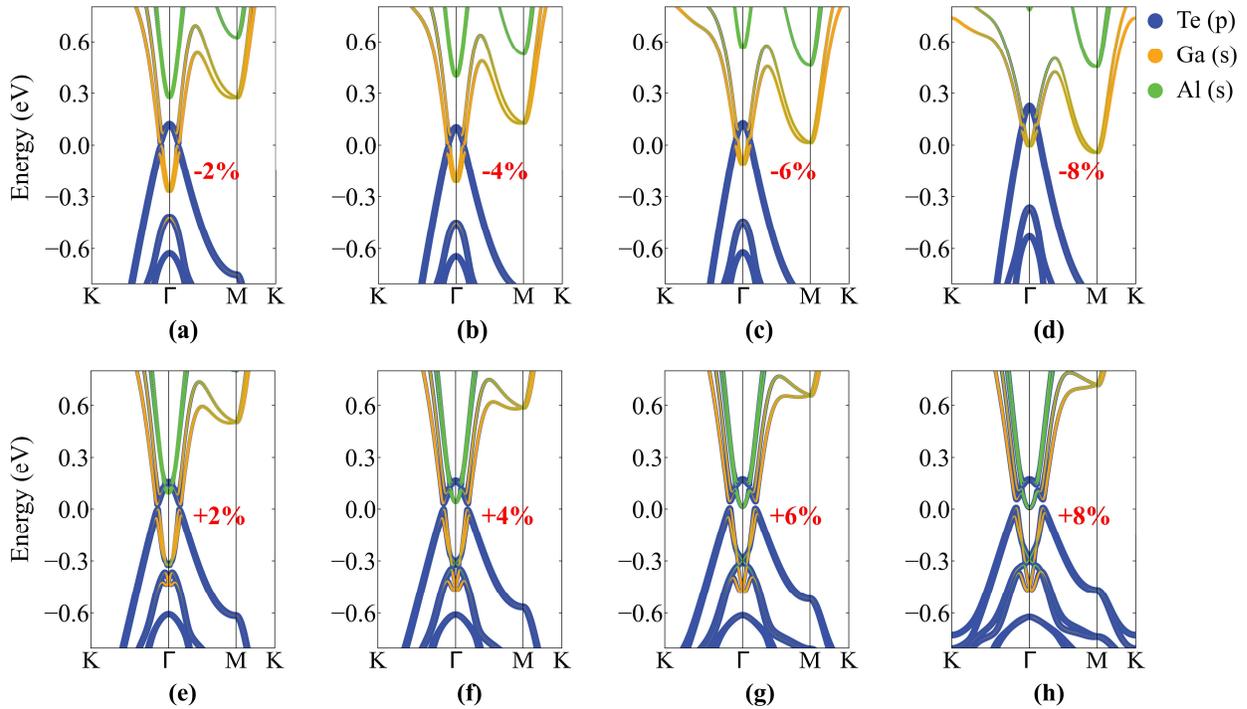

**Fig. 4. Band structures of SrAlGaTe$_4$ under various strain conditions with included SOC.** Panels (**a-d**) display results for -2% to -8% biaxial strain, while panels (**e-h**) represent the results for +2% to +8% strain.

We have also conducted a thorough investigation into the impact of strain on SrAlGaTe$_4$, as depicted in **Fig. 4**. Notably, at -8% and -6% strain, the CBM resides in the M point, and there is no band gap, resulting in a metallic state. Moving on to -4% strain, the CBM relocates close to the Γ point (but not right at the Γ point), leading to an inverted bandgap and the QSH state. The bandgap expands and the band inversion persists at +2% and +4% strain, as evidenced in **Figs. 4(e)** and **4(f)**. However, as we increase the strain to +6%, the CBM precisely locates to the Γ point, resulting in a transition from a QSH insulator to a metallic phase due to the bandgap closure. This metallic phase persists as the strain is further increased to +8%. Furthermore, we have observed a similar transition from the QSH state to a metallic state in CaAlGaTe$_4$. The relevant band structure data for CaAlGaTe$_4$ is provided in section F of the **ESI**, and the corresponding band structures without SOC can be found in section G of the **ESI**. Such strain-tunable transitions have also been observed in Bi$_2$Se$_3$,[29] 1S-MoSe$_2$, MoTe$_2$, and WS$_2$.[28] We summarized the strain dependent phases of MAlGaTe$_4$ (M: Mg, Sr) in **Fig. 5**.

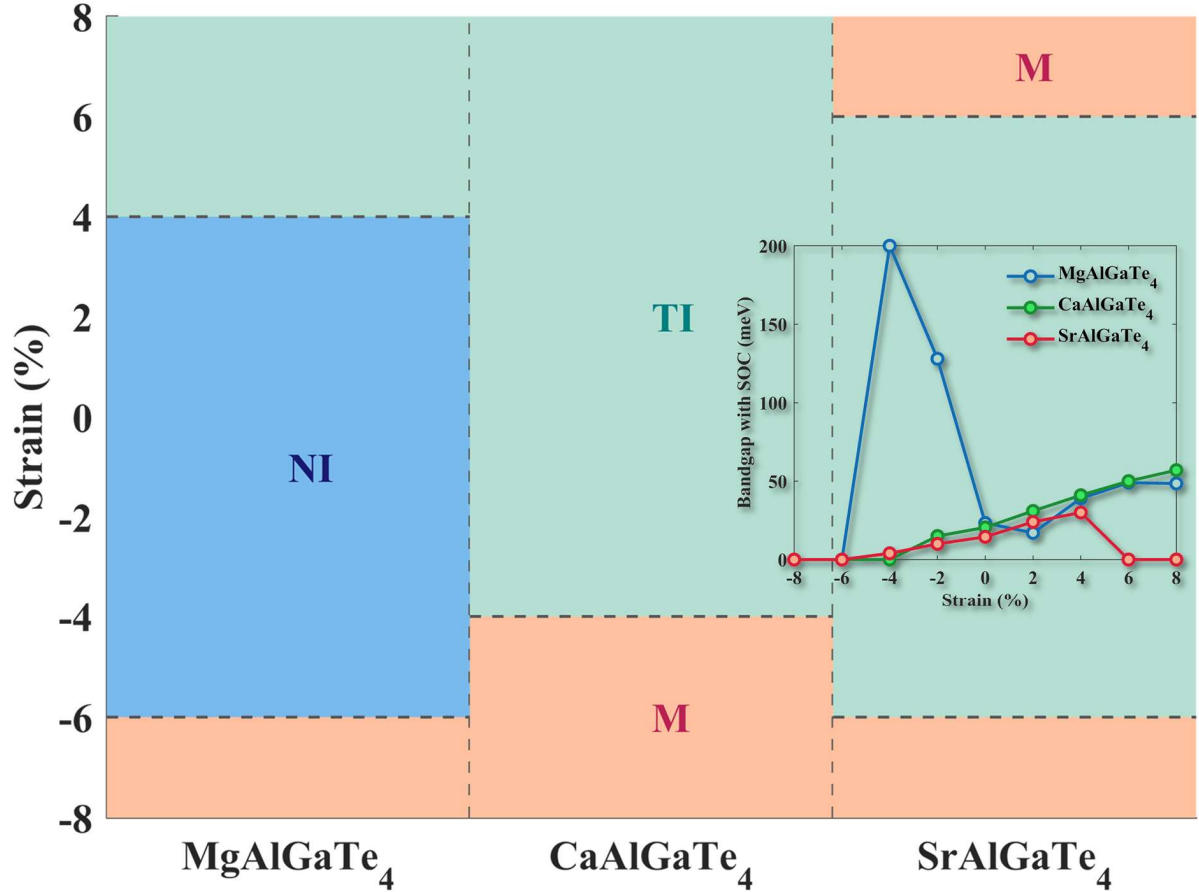

**Fig. 5. Phase diagram of (Ca, Sr, Mg)AlGaTe₄ as a function of applied biaxial strain.** The inset illustrates the evolution of their SOC-included bandgaps with strain. Here, M = metal, NI = normal/trivial insulator, and TI = 2D topological insulator or QSH insulator.

Before concluding, it is worth noting that, as extensively detailed in section H of the **ESI**, MAlGaTe$_4$ (M: Mg, Ca, Sr) exhibit notable in-plane and modest out-of-plane piezoelectric properties. This unique characteristic positions these materials at the intersection of strain-tunable electronic and topological properties, coupled with piezoelectricity. This combination can be used for innovative applications in the future.

**Conclusion**

In this study, we introduce a novel family of monolayer piezoelectric materials, MAlGaTe$_4$ (M: Mg, Ca, Sr). These materials exhibit intriguing phase transitions, shifting between metal, normal/trivial insulator, and topological insulator phases under the influence of applied strain. Among them, MgAlGaTe$_4$ stands out as particularly fascinating. Initially a normal insulator at zero strain, it transitions into either the QSH phase or a metallic state, contingent on the specific strain applied. In contrast, CaAlGaTe$_4$ and SrAlGaTe$_4$ already possess the QSH phase in their unstrained state. Remarkably, the application of biaxial strain to these compounds induces a series of phase transitions, spanning metallic and topological insulator phases. Thus, our work introduces a material platform where both metal-insulator transition and topological phase transitions can be reversibly controlled via strain. This holds promising implications for the advancement of next-generation, low-power topological devices.

**Computational Details:**

**Electronic Structure Calculations:** First-principles calculations based on density functional theory (DFT)[30,31] were performed using the Vienna Ab initio Simulation Package (VASP).[32–36] The projector-augmented wave (PAW) method[37] and a plane-wave basis set were employed. The Perdew–Burke–Ernzerhof (PBE) functional within generalized gradient approximation (GGA) was used for exchange-correlation energy.[38] In case of geometry optimization, atoms and cell parameters were relaxed until residual forces were smaller than 0.002 eV/Å. An energy convergence threshold of $10^{-8}$ eV and an energy cutoff of 450 eV (400 eV for SrAlGaTe$_4$) were set. A 16 × 16 × 1 Monkhorst-pack k-point grids were used for Brillouin zone sampling.

**Strained Monolayers:** For strained monolayers, force and energy convergence thresholds were set at 0.01 eV/Å and $10^{-5}$ eV respectively, with an energy cutoff of 315 eV. A $16 \times 16 \times 1$ gamma-centered k-mesh was used.

**Visualization and Plotting:** Crystal structures were visualized using VESTA[39] and XCrySDen[40]. Figures were generated using MATLAB, Gnuplot,[41] and Sumo.[42]

**Topological Invariant Analysis:** Wannier90[43] and WannierTools[44] were used to calculate the topological invariant and edge density of states. Maximally Localized Wannier functions (MLWF) were employed to construct an MLWF tight-binding Hamiltonian.

**Edge State Spectrum:** The edge state spectrum was computed using the iterative Green's function method[45] implemented in WannierTools, in conjunction with Quantum ESPRESSO (QE).[46–49] Norm-conserving Vanderbilt pseudopotentials[50] were used in QE. Prior to further calculations, structures were re-relaxed in QE. A kinetic energy cutoff of 85 Ry was employed for wavefunction.

**Piezoelectric Properties:** Density functional perturbation theory (DFPT) with GGA+SOC in VASP was used to calculate piezoelectric properties. A $5 \times 10 \times 1$ gamma-centered k-mesh was employed.

**Phonon Bandstructures:** Phonon bandstructures were calculated using DFPT, as implemented in CASTEP.[51]

**Acknowledgement:**

Financial grant to procure VASP is given by the Bangladesh University of Engineering and Technology (BUET). Computational facilities provided by BUET and Princeton University are duly acknowledged. Works at Princeton University are supported by the Gordon and Betty Moore Foundation (GMBF4547 and GMBF9461).

**Supporting Information Available:**

A. Dynamic Stability, B. Mechanical Properties, C. Crystal Structure and Orbital-Resolved Bandstructures of CaAlGaTe$_4$, D. Edge Density of States and $\mathbb{Z}_2$ for CaAlGaTe$_4$, E. Orbital-Resolved Bandstructures of MgAlGaTe$_4$ at Certain Strain Values, F. Orbital-Resolved Bandstructures of CaAlGaTe$_4$ with Biaxial Strain (SOC Included), G. Orbital-Resolved Bandstructures of CaAlGaTe$_4$ and SrAlGaTe$_4$ with Biaxial Strain (without SOC), H. Piezoelectric Properties, References.

**Author Information:**

**Corresponding Authors:**

Md. Shafayat Hossain[1], M. Zahid Hasan[2], Quazi D. M. Khosru[3]

[1, 2]Department of Physics, Princeton University, Princeton, NJ, 08544, USA

[3]Department of Electrical and Electronic Engineering, Bangladesh University of Engineering and Technology, Dhaka, 1205, Bangladesh

E-mails: [1]mdsh@princeton.edu; [2]mzhasan@princeton.edu; [3]qdmkhosru@eee.buet.ac.bd.

**Author Contributions:**

MSH and QDMK conceived the project. TTT contributed to ideation, simulation, analysis of results, manuscript writing, and editing. MSH contributed to ideation, analysis of results, manuscript review, and editing. NTH performed simulation, analysis of results, and manuscript writing. AR contributed to software familiarization, and manuscript writing. MZH and QDMK reviewed and edited the whole manuscript. The project has been supervised by MSH, MZH, and QDMK.

**Competing Interests:**

The authors declare no competing financial or non-financial interests.

**Electronic Supplementary Information (ESI)**

# Tunable Topological Phase Transitions in a Piezoelectric Janus Monolayer

**A. Dynamic Stability**

To assess the dynamic stability of the MAlGaTe$_4$ family of materials (where M represents Mg, Ca, or Sr), we conducted a thorough examination of their phonon dispersion spectra, which are presented in **Fig. S1**. Notably, the phonon spectra do not contain any negative frequency components or imaginary modes of vibration, indicating a robust dynamic stability. Within each phonon dispersion spectrum, we observe a total of 21 distinct modes of vibration, owing to the seven-atom composition of each unit cell. Among these, three modes are categorized as acoustic modes (demonstrating zero frequency at the Γ point), while the remaining 18 modes are designated as optical modes. For MgAlGaTe$_4$, the optical spectrum encompasses five doubly degenerate modes alongside eight non-degenerate modes. Conversely, in the case of CaAlGaTe$_4$ and SrAlGaTe$_4$, there are four doubly degenerate modes and ten non-degenerate modes. It is worth mentioning that the frequencies of the optical modes exhibit notable similarity across all three monolayers. In particular, the frequency of the optical modes is nearly equal for the three monolayer compounds. Furthermore, due to the overlap of acoustic and optical modes at certain frequencies, strong optical-acoustic scattering might exist in these materials. This phenomenon, in turn, suggests the likelihood of low thermal conductivity.[1]

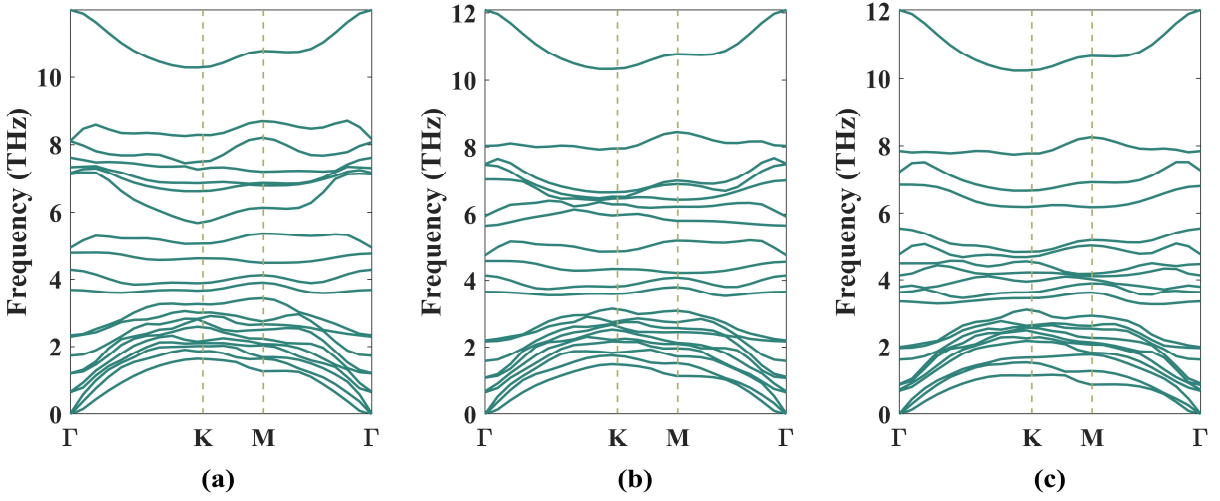

**Fig. S1.** Phonon dispersion spectra of monolayer **(a)** MgAlGaTe$_4$, **(b)** CaAlGaTe$_4$, and **(c)** SrAlGaTe$_4$.

## B. Mechanical Properties

After confirming the dynamic stability of the monolayers, we proceeded to examine the mechanical properties of MAlGaTe$_4$. To achieve this, we computed the elastic constants tensor ($C_{kj}$) for each material. Due to the hexagonal lattice structure, $C_{11} = C_{22}$. Moreover, $C_{66}$ can be derived using the expression $C_{66} = \frac{C_{11}-C_{12}}{2}$. Therefore, there are only two independent elastic constants, $C_{11}$ and $C_{12}$. To adapt these constants for 2D materials, we multiplied the raw values (in $Nm^{-2}$) with the out-of-plane lattice constant, c. The two-dimensional Young's modulus, $Y_{2D}$ and Poisson's ratio, $\nu$ were determined using the following expressions:[2] $Y_{2D} = \frac{C_{11}^2 - C_{12}^2}{C_{11}}$ and $\nu = \frac{C_{12}}{C_{11}}$.

**Table S1** provides a comprehensive list of the obtained elastic constants, along with the calculated two-dimensional Young's moduli and Poisson's ratios.

**Table S1.** Elastic constants ($C_{kj}$), Young's moduli ($Y_{2D}$) and Poisson's ratios ($v$) of (Ca, Sr, Mg)AlGaTe$_4$ monolayers.

| Material | $C_{11}(Nm^{-1})$ | $C_{12}(Nm^{-1})$ | $C_{66}(Nm^{-1})$ | $Y_{2D}(Nm^{-1})$ | $v$ |
|---|---|---|---|---|---|
| MgAlGaTe$_4$ | 84.02 | 28.07 | 27.98 | 74.65 | 0.334 |
| CaAlGaTe$_4$ | 69.82 | 25.60 | 22.11 | 60.44 | 0.367 |
| SrAlGaTe$_4$ | 64.50 | 24.30 | 20.10 | 55.35 | 0.377 |

For each material, $C_{11} > 0$ and $C_{11}^2 - C_{12}^2 > 0$. This ensures that they meet Born's criteria for mechanical stability.[2,3] Moreover, given that $v > \frac{1}{3}$, it can be concluded that all the monolayers exhibit ductility in accordance with the Frantsevich rule.[4,5]

The Young's modulus of MgAlGaTe$_4$ ($74.65\ Nm^{-1}$) is comparable to that of SrAlGaSe$_4$ ($71.68\ Nm^{-1}$).[6] Meanwhile, the Young's moduli of CaAlGaTe$_4$ ($60.44\ Nm^{-1}$) and SrAlGaTe$_4$ ($55.35\ Nm^{-1}$) fall within the range of 2D materials like GeSO ($59.90\ Nm^{-1}$),[2] SnSSe ($57.50\ Nm^{-1}$),[7] PdSSe ($62.33\ Nm^{-1}$ along x, $64.68\ Nm^{-1}$ along y),[8] etc. Nevertheless, they are all notably lower than those of MoS$_2$ ($130.0\ Nm^{-1}$),[9] MoSSe ($113.0\ Nm^{-1}$),[10] 2D black phosphorene ($214 \pm 11\ Nm^{-1}$),[11] Al$_2$SeO ($94.0\ Nm^{-1}$),[4] Al$_2$SO ($111.3\ Nm^{-1}$),[4] MA$_2$Z$_4$ (M = Mo, W; A = Si, Ge and Z = N, P, As),[12] etc. This lower $Y_{2D}$ value implies reduced in-plane stiffness, leading to the conclusion that none of the proposed materials exhibit high stiffness. The elastic constants will be revisited in a later section when discussing the piezoelectric properties.

## C. Crystal Structure and Orbital-Resolved Band structures of CaAlGaTe$_4$

The lattice constants and bond-lengths relevant to CaAlGaTe$_4$ are provided in **Table S2**. In the absence of SOC (as illustrated in **Fig. S2(b)**), the band structure of CaAlGaTe$_4$ is characterized by a gapless nature. Upon the inclusion of SOC (as presented in **Fig. S2(c)**), this results in the emergence of an inverted bandgap with a magnitude of 20.5 meV. Akin to the behavior observed in SrAlGaTe$_4$, the signature of topological non-triviality marked by electronic band inversion is also discernible in CaAlGaTe$_4$.

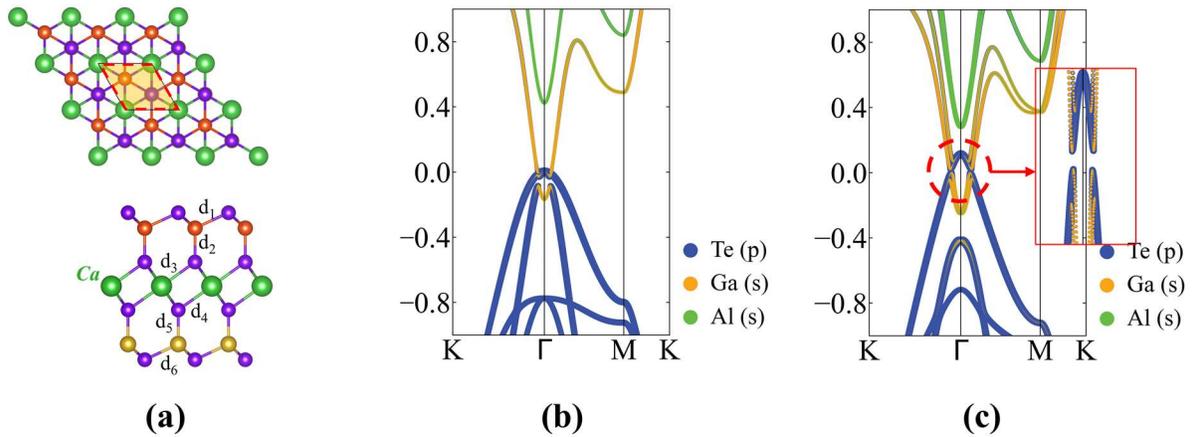

**Fig. S2.** (a) Crystal structure of CaAlGaTe$_4$, depicting the top-view (above) and side-view (below). (b) and (c) show the orbital-resolved band structure of CaAlGaTe$_4$ without and with SOC, respectively.

**Table S2.** Lattice constants and bond-lengths of CaAlGaTe$_4$.

| Material | Lattice constant (Å) | $d_1$ (Å) | $d_2$ (Å) | $d_3$ (Å) | $d_4$ (Å) | $d_5$ (Å) | $d_6$ (Å) |
|---|---|---|---|---|---|---|---|
| CaAlGaTe$_4$ | 4.35 | 2.76 | 2.53 | 3.09 | 3.08 | 2.55 | 2.77 |

## D. Edge Density of States and $\mathbb{Z}_2$ for CaAlGaTe$_4$

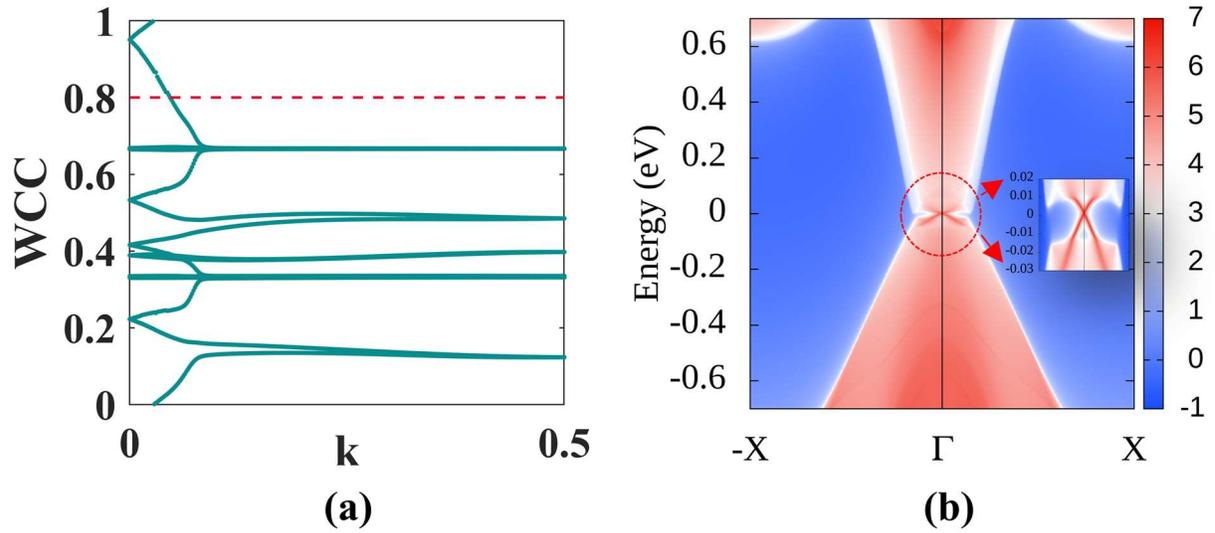

**Fig. S3.** Topological invariants and edge modes for unstrained monolayer CaAlGaTe$_4$. **(a)** WCC evolution curves and **(b)** edge density of states displaying gapless edge states. The WCC curves intersect the horizontal reference line once (odd), indicating $\mathbb{Z}_2 = 1$.

## E. Orbital-Resolved Band Structures of MgAlGaTe$_4$ at Certain Strain Values

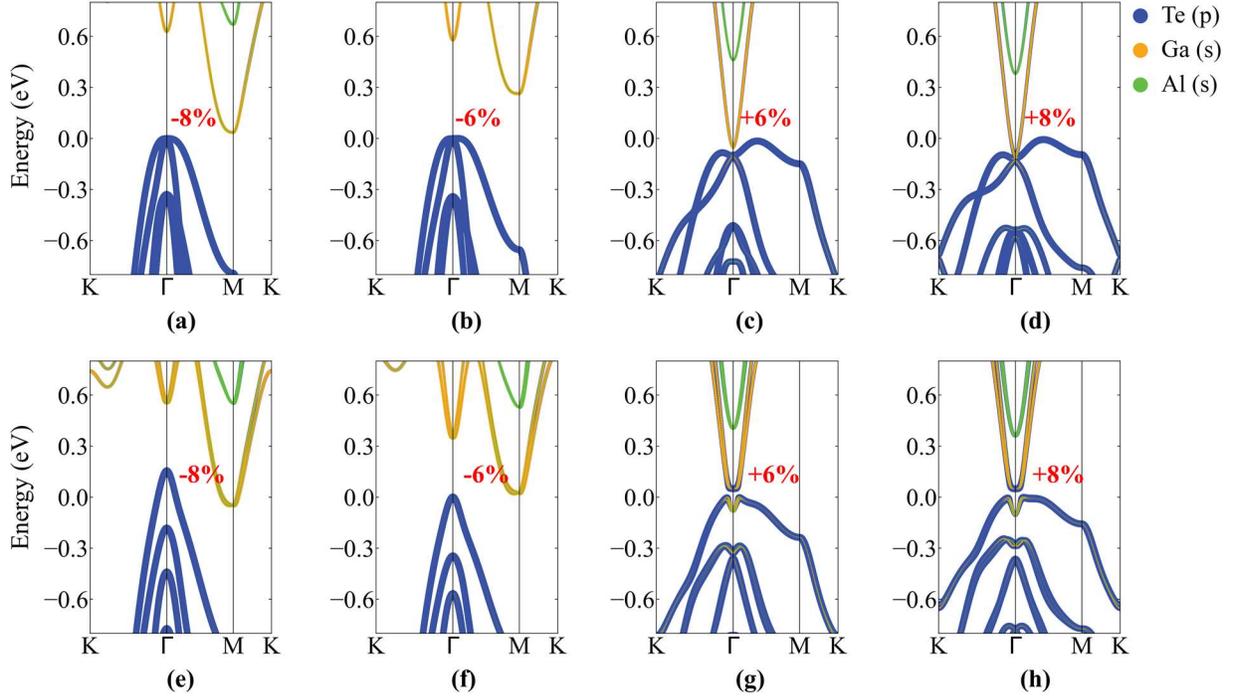

**Fig. S4.** Orbital-resolved band structures of monolayer MgAlGaTe$_4$ under **(a, e)** -8%, **(b, f)** -6%, **(c, g)** +6%, and **(d, h)** +8% strain. Panels **(a-d)** display band structures without SOC, while panels **(e-h)** show the corresponding band structures with SOC.

## F. Orbital-Resolved Band Structures of CaAlGaTe$_4$ with Biaxial Strain (SOC Included)

We have conducted a thorough investigation into the impact of strain on CaAlGaTe$_4$, as depicted in **Fig. S5**. Notably, at -8%, -6%, and -4% strain, the CBM resides in the M point, and there is no band gap, resulting in a metallic state. Moving on to -2% strain, the CBM relocates to the Γ point, leading to an inverted bandgap and the QSH state. As the strain changes from -2% to 8%, the bandgap gradually expands and the band inversion persists, as evidenced in **Figs. 4(e)** and **4(f)**.

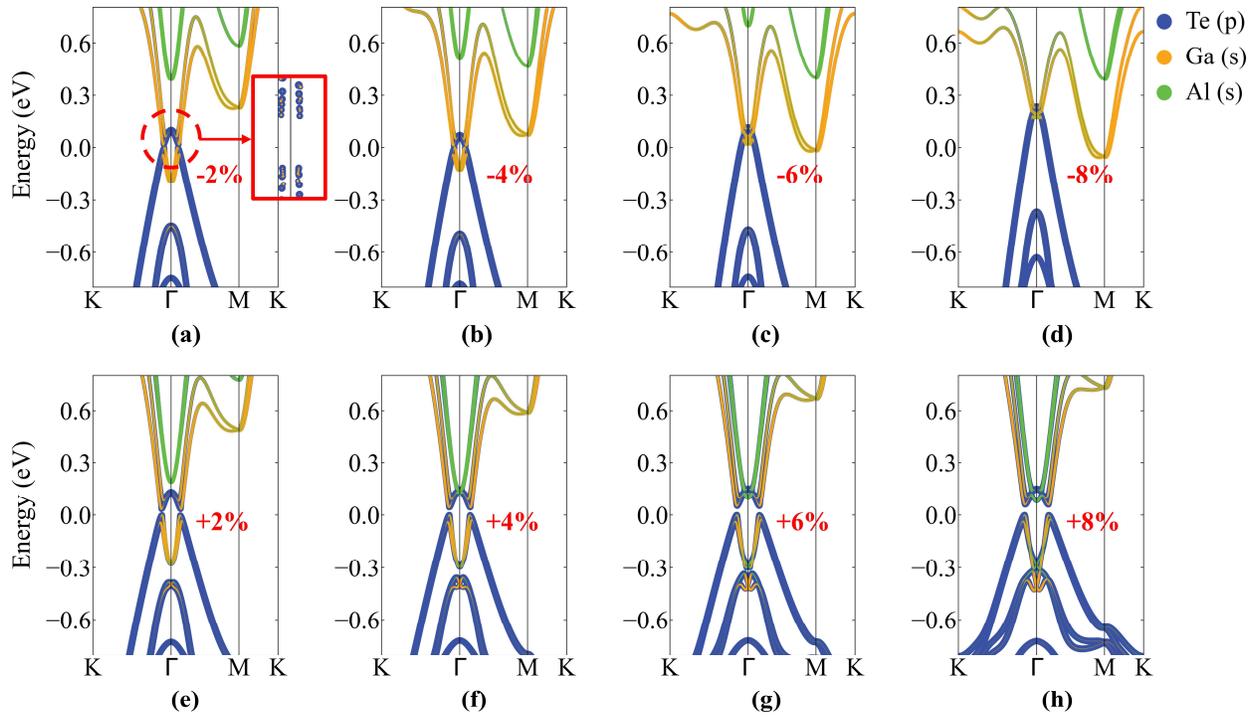

**Fig. S5. Band structures of CaAlGaTe$_4$ under various strain conditions with included SOC.** Panels (**a-d**) display results for -2% to -8% biaxial strain, while panels (**e-h**) represent the results for +2% to +8% strain.

## G. Orbital-Resolved Band Structures of CaAlGaTe$_4$ and SrAlGaTe$_4$ with Biaxial Strain (without SOC)

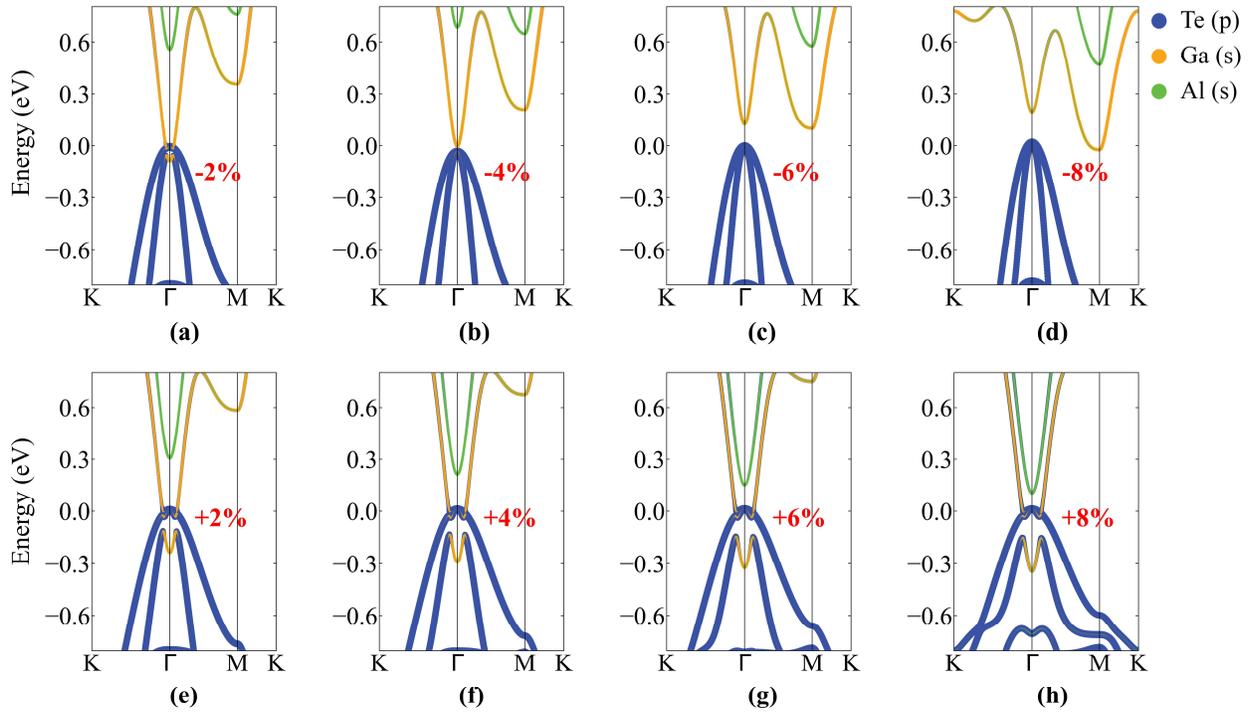

**Fig. S6. Band structures of CaAlGaTe$_4$ under various strain conditions without considering SOC.** Panels (**a-d**) display results for -2% to -8% biaxial strain, while panels (**e-h**) represent the results for +2% to +8% strain.

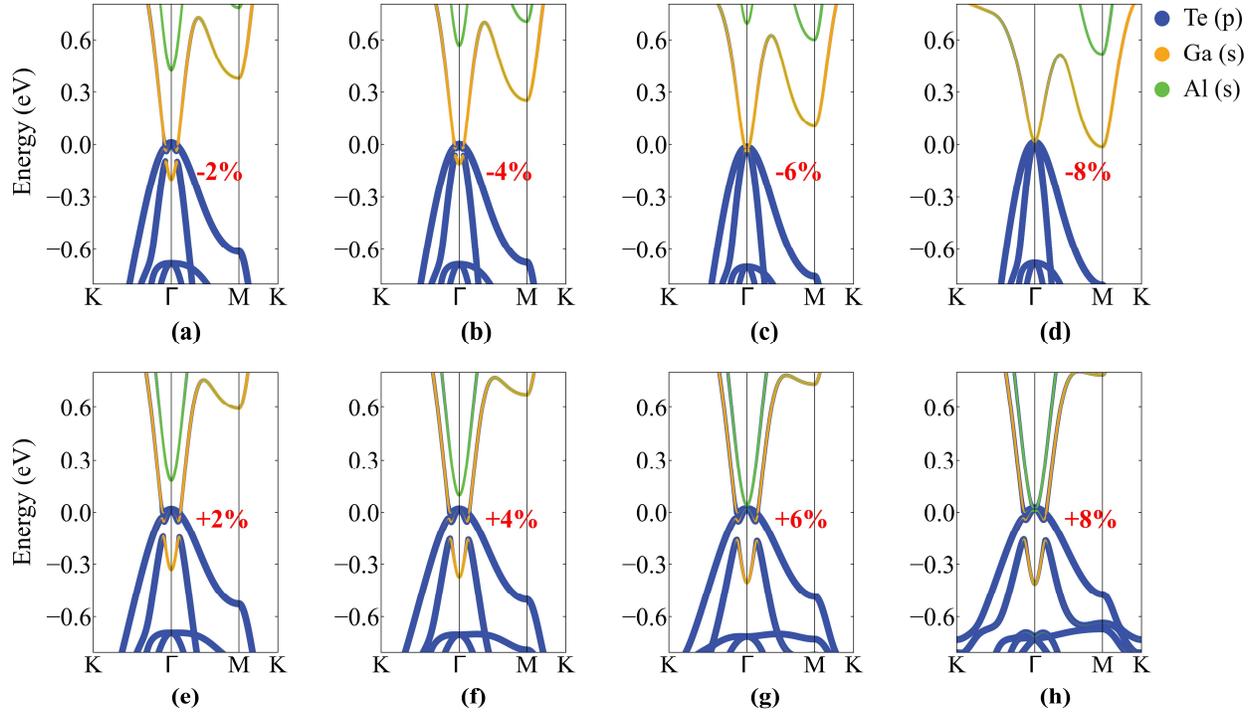

**Fig. S7. Band structures of SrAlGaTe4 under various strain conditions without considering SOC**. Panels (a-d) display results for -2% to -8% biaxial strain, while panels (e-h) represent the results for +2% to +8% strain.

## H. Piezoelectric Properties

In this section, we investigate the piezoelectric properties of the MAlGaTe$_4$ family of materials, where M represents Mg, Ca, or Sr. Prior studies have demonstrated piezoelectricity in various Janus materials.[13–15] Furthermore, the coexistence of piezoelectricity and the QSH effect has been reported for SrAlGaSe$_4$.[6] Materials exhibiting both these exotic properties hold significant potential for the design of advanced, high-performance devices. To explore the piezoelectric properties, we selected MgAlGaTe$_4$ at 4% strain, unstrained CaAlGaTe$_4$, and unstrained SrAlGaTe$_4$ as the materials of interest. An orthorhombic supercell was employed as the computational unit to calculate the piezoelectric stress coefficients ($e_{ij}$) using DFPT with GGA+SOC. The supercell corresponding to MgAlGaTe$_4$ at 4% strain is depicted in the inset of **Fig. S8(a)**. The piezoelectric stress coefficients, re-normalized by multiplying with the lattice constant, $c$ to convert them into appropriate units for 2D materials, are composed of ionic and electronic contributions. The piezoelectric stress and strain ($d_{kj}$) coefficients are related by the equation:

$$e_{ij} = d_{ik} C_{kj} \tag{1}$$

Given their 3m point-group symmetry, $C_{kj}$, $e_{ij}$, and $d_{ik}$ for MgAlGaTe$_4$ can be expressed as:[16]

$$C_{kj} = \begin{pmatrix} C_{11} & C_{12} & 0 \\ C_{12} & C_{11} & 0 \\ 0 & 0 & \frac{C_{11}-C_{12}}{2} \end{pmatrix} \tag{2}$$

$$e_{ij} = \begin{pmatrix} e_{11} & -e_{11} & 0 \\ 0 & 0 & -e_{11} \\ e_{31} & e_{31} & 0 \end{pmatrix} \tag{3}$$

$$d_{ik} = \begin{pmatrix} d_{11} & -d_{11} & 0 \\ 0 & 0 & -2d_{11} \\ d_{31} & d_{31} & 0 \end{pmatrix} \tag{4}$$

By utilizing the equations (1-4), $d_{11}$ and $d_{31}$ can be expressed as:

$$d_{11} = \frac{e_{11}}{C_{11}-C_{12}} \tag{5}$$

$$d_{31} = \frac{e_{31}}{C_{11}+C_{12}} \tag{6}$$

The computed $e_{ij}$, comprising ionic and electronic contributions, and $d_{ik}$ for the three candidate materials are listed in **Table S3**. Additionally, for visualization purposes, the values are plotted in **Fig. S8**.

From **Fig. S8(a)**, it is evident that $e_{11}$ is predominantly composed of electronic contributions for each monolayer. Among the MAlGaTe$_4$ family of materials, MgAlGaTe$_4$ exhibits the highest absolute value of $e_{11}$, followed by CaAlGaTe₄ and SrAlGaTe₄, in that order. On the other hand, the values of $e_{31}$ do not exhibit a systematic trend. The ionic and electronic contributions to $e_{31}$ are of opposite sign, resulting in a net value that is non-zero but not particularly large, in each case. For Mg and Sr, the ionic contribution dominates, determining the sign of $e_{31}$. However, in the case of Ca, the opposite phenomenon occurs. **Fig. S8(b)-(d)** further confirms that $e_{31}$, $d_{11}$, and $d_{31}$ do not exhibit any discernible increasing or decreasing trend as one progresses from Mg to Ca to Sr. The in-plane piezoelectric strain coefficient, $d_{11}$ has a value of 8.25 pm/V for MgAlGaTe₄, which is comparable to that of BiTeI (8.20 pm/V),[16] Bi₂Te₂Se (8.34 pm/V),[14] and CaAlGaSe₄ at 6% tensile strain (8.07 pm/V).[6] The values for CaAlGaTe₄ and SrAlGaTe₄ are -1.99 pm/V and -1.37 pm/V, respectively, which are comparable to those of WSSe (2.02 pm/V)[17] and NiClBr (1.35 pm/V),[18] respectively. The values of out-of-plane piezoelectric strain coefficients ($d_{31}$) are 0.05 pm/V, -0.004 pm/V, and 0.02 pm/V for MgAlGaTe₄ at 4% tensile strain, unstrained CaAlGaTe₄, and unstrained SrAlGaTe₄, respectively. These values of $d_{31}$ are relatively small for all the three materials. However, they are still comparable to those of other reported piezoelectric monolayers, such as, Bi₂Se₂Te (0.06 pm/V),[14] Bi₂Te₂S (0.03 pm/V),[14] Bi₂Te₂Se (0.01 pm/V),[14] MoSSe (0.02 pm/V),[17] MoSTe (0.028 pm/V),[16] BiTeF (0.07 pm/V),[19] WSTe (0.007 pm/V),[17] and WSeTe (0.008 pm/V).[17] These findings suggest that monolayer MAlGaTe₄ holds promise for applications in piezoelectric devices.

**Table S3.** Elastic constants, $C_{kj}$ (in $Nm^{-1}$), piezoelectric stress coefficients, $e_{ij}$ ($10^{-10}\ C/m$), and piezoelectric strain coefficients, $d_{ik}$ ($pm/V$) of (Ca, Sr, Mg)AlGaTe₄ monolayers. The table includes the ionic (Ion.), electronic (Elec.), and total (Tot.) contributions to $e_{ij}$.

| Material | $C_{11}$ | $C_{12}$ | $e_{11}$ | | | $e_{31}$ | | | $d_{11}$ | $d_{31}$ |
|---|---|---|---|---|---|---|---|---|---|---|
| | | | Ion. | Elec. | Tot. | Ion. | Elec. | Tot. | | |
| MgAlGaTe$_4$ (at 4%) | 66.47 | 21.92 | 0.114 | 3.56 | 3.674 | 0.061 | -0.017 | 0.045 | 8.25 | 0.05 |
| CaAlGaTe$_4$ | 69.82 | 25.60 | -0.340 | -0.538 | -0.878 | 0.061 | -0.064 | -0.004 | -1.99 | -0.004 |
| SrAlGaTe$_4$ | 64.50 | 24.30 | -0.165 | -0.385 | -0.550 | 0.039 | -0.021 | 0.017 | -1.37 | 0.02 |

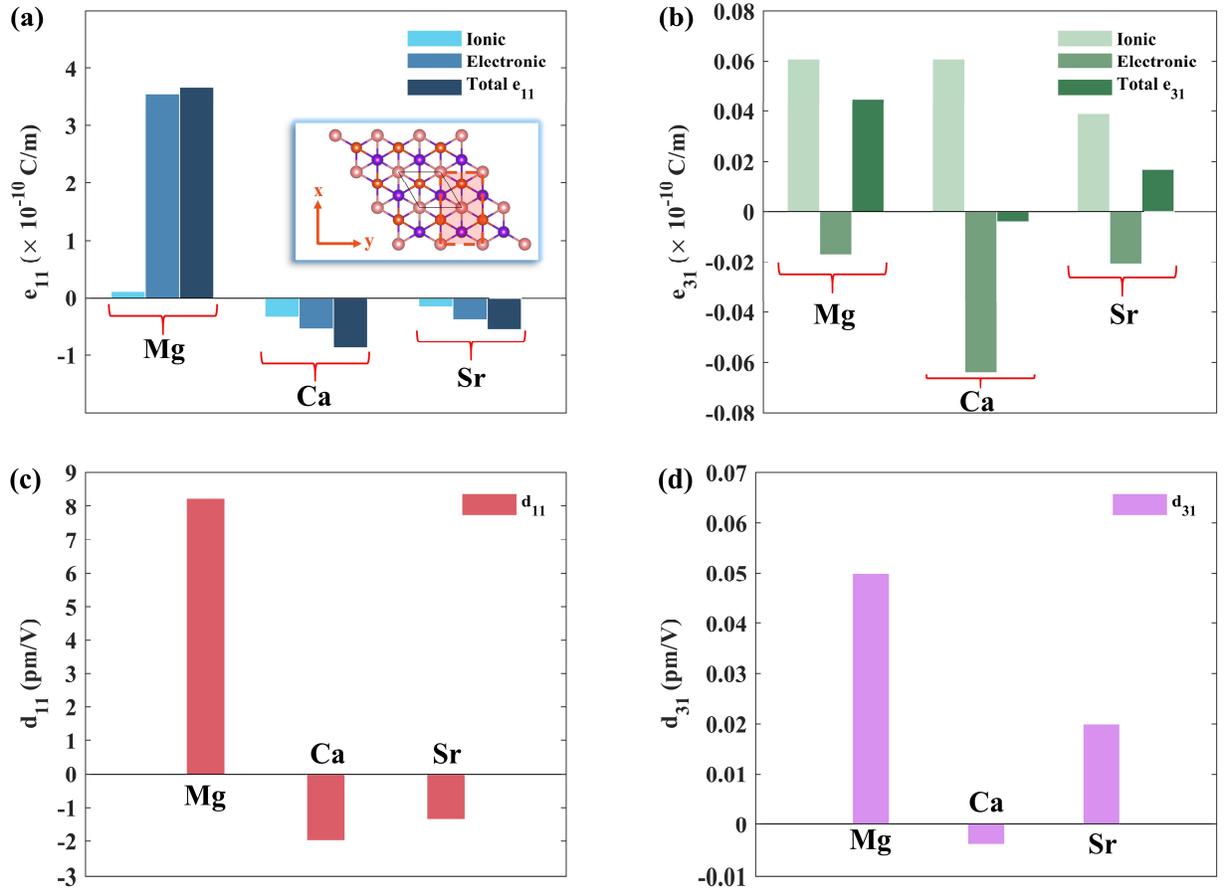

**Fig. S8.** Bar charts illustrating the ionic and electronic contributions to the piezoelectric stress coefficients, **(a)** $e_{11}$ and **(b)** $e_{31}$, along with their total values. The semi-transparent red rectangle, as shown in the inset of **(a)**, represents the orthorhombic supercell used for calculation. Charts in **(c)** and **(d)** represent the piezoelectric strain coefficients, $d_{11}$ and $d_{31}$, respectively. In these charts,

'Mg,' 'Ca,' and 'Sr' correspond to MgAlGaTe$_4$ at 4% tensile strain, unstrained CaAlGaTe$_4$, and unstrained SrAlGaTe$_4$, respectively.